\documentclass[conference,final]{IEEEtran}

\usepackage{acronym}
\usepackage{amsfonts}
\usepackage[dvips]{graphicx}
\usepackage{times}
\usepackage{cite}
\usepackage{amsmath}
\usepackage{array}
\usepackage{amssymb}
\usepackage{stfloats}
\usepackage{slashbox}
\usepackage{graphicx}
\usepackage{footnote}
\usepackage{amsthm}
\usepackage{booktabs}
\usepackage{array}
\usepackage{algorithmic}
\usepackage{algorithm}
\usepackage{subeqnarray}
\usepackage{cases}
\usepackage{threeparttable}
\usepackage{color}
\usepackage{epstopdf}
\usepackage{multirow}
\usepackage{tabularx}
\usepackage{enumerate}
\usepackage{multicol}

\newcommand{\figref}[1]{{Fig.}~\ref{#1}}

\def\bb0{{\mathbb{0}}}

\def\bb{{\mathbf{b}}}

\def\bff{{\mathbf{f}}}

\def\bh{{\mathbf{h}}}

\def\bv{{\mathbf{v}}}

\def\bx{{\mathbf{x}}}

\def\b0{{\mathbf{0}}}

\def\sf0{{\mathsf{0}}}

\def\rm0{{\mathrm{0}}}

\def\Nt{{N_\mathrm{t}}}

\def\Pt{{P_{\mr{t}}}}
\def\Hb{{\mathcal{H}_{\mr{b}}}}

\newcommand{\mb}{\mathbf}
\newcommand{\mr}{\mathrm}
\def\j{\mathrm{j}}

\acrodef{CSI}[CSI]{channel state information}
\acrodef{CSIT}[CSIT]{channel state information at the transmitter}
\acrodef{CSIR}[CSIR]{channel state information at the receiver}
\acrodef{MIMO}[MIMO]{multiple-input multiple-output}
\acrodef{SISO}[SISO]{single-input single-output}
\acrodef{MISO}[MISO]{multiple-input single-output}
\acrodef{SIMO}[SIMO]{single-input multiple-output}
\acrodef{ADCs}[ADCs]{analog-to-digital convertors}
\acrodef{SNR}[SNR]{signal-to-noise ratio}
\acrodef{AWGN}[AWGN]{additive white Gaussian noise}
\acrodef{MRT}[MRT]{maximal ratio transmission}

\IEEEoverridecommandlockouts
\begin{document}
%

\title{Limited Feedback in Multiple-Antenna Systems with One-Bit Quantization}
\author{
	\authorblockN{Jianhua~Mo and Robert W. Heath Jr.}
	\IEEEauthorblockA{Wireless Networking and Communications Group\\
	The University of Texas at Austin, Austin, TX 78712, USA}
	Email: \{jhmo, rheath\}@utexas.edu
	\thanks{This material is based upon work supported in part by the National Science Foundation under Grant No. NSF-CCF-1527079.}
}
\maketitle
\begin{abstract} 
Communication systems with low-resolution analog-to-digital-converters (ADCs) can exploit channel state information at the transmitter (CSIT) and receiver. This paper presents initial results on codebook design and performance analysis for limited feedback systems with one-bit ADCs. Different from the high-resolution case, the absolute phase at the receiver is important to align the phase of the received signals when the received signal is sliced by one-bit ADCs. A new codebook design for the beamforming case is proposed that separately quantizes the channel direction and the residual phase.
\end{abstract}

\section{Introduction}

The large bandwidth of the emerging wireless network, for example, millimeter wave WLAN \cite{Baykas_COMM11} or 5G millimeter wave cellular \cite{Rappaport_TCOM15,Heath_JSTSP15, Rappaport_Book15}, has introduced new challenges for the hardware design. A main challenge is the power assumption with analog-to-digital converters (ADCs). Data summarized in \cite{Murmann_15} show that ADC power consumption increases drastically for sampling frequency above $100$ megasamples per second.
A direct solution to this bottleneck is to use power efficient low-resolution -- even one-bit -- ADCs.

The use of few- and especially one-bit ADCs radically changes both the theory and practice of communication, for example, the capacity analysis \cite{Dabeer_SPAWC06, Singh_TCOM09, Mezghani_ISIT07, Mezghani_ISIT12, Bai_Qing_ETT15, Mo_Jianhua_TSP15},  channel estimation \cite{Ivrlac_WSA07, Dabeer_ICC10, Mezghani_WSA10, Mezghani_WSA12, Mo_Jianhua_Asilomar14, Jacobsson_arxiv15, Wen_Chao-Kai_arxiv15b}, and symbol detection \cite{Mezghani_MELECON12, Kamilov_TSP12, Wang_Shengchu_TWC15}. At present, the exact capacity of quantized \ac{MIMO} channel is generally unknown, except for the simple \ac{MISO} channel and some special cases, such as in the low or high SNR regime \cite{Dabeer_SPAWC06, Mezghani_ISIT07, Singh_TCOM09, Mo_Jianhua_TSP15}. Transmitting independent QPSK signals \cite{Mezghani_ISIT07} or Gaussian signals \cite{Mezghani_ISIT12, Bai_Qing_ETT15} from each antenna nearly achieves the capacity at low SNR, but is far from the optimal at high SNR.
In our work \cite{Mo_Jianhua_TSP15}, where \ac{CSIT} is assumed, simple channel inversion precoding (versus the usual eigenbeamforming) is nearly optimal if the channel has full row rank. If the channel is low rank, we proposed a new precoding method achieving the capacity at high SNR. MIMO precoding provides a substantial performance improvement compared with the no-precoding case (i.e., independent QPSK or Gaussian signaling).
%
%
%

Despite the potential gain of precoding and therefore the importance of \ac{CSIT}, there is -- to our knowledge -- no work that has considered limited feedback with low-resolution ADCs. The results on limited feedback with infinite-resolution ADCs, for example, \cite{Love_IT03, Jindal_IT06, Love_JSAC08}, cannot be directly extended to the low-resolution ADCs -- the fundamentals are different. For example, in MISO limited feedback beamforming \cite{Love_IT03}, the optimum beamformer $\bff$ is phase invariant, meaning that equivalent performance is also achieved by $\bff e^{j \theta}$. In our capacity results in \cite{Mo_Jianhua_TSP15}, we found that the optimum beamformer was the matched filter, and in fact was not phase invariant. The reason is that phase at the receiver is important when the received signal is sliced by 1-bit ADCs; an important function of \ac{CSIT} is to align the phase of the received signals. One implication is that phase-invariant Grassmannian beamforming codebooks \cite{Love_IT03} will no longer be appropriate.

In this paper, we develop limited feedback schemes for \ac{SISO} and \ac{MISO} systems with one-bit ADCs. Our approach leverages recent work \cite{Ivrlac_WSA07, Dabeer_ICC10, Mezghani_WSA10, Mezghani_WSA12, Mo_Jianhua_Asilomar14, Jacobsson_arxiv15, Wen_Chao-Kai_arxiv15b} that shows that it is possible to estimate the MIMO channel even with one-bit ADCs at the receiver.
Given an estimate of the channel, we propose a new codebook design that explicitly incorporates the phase sensitivity required for feedback in the one-bit ADC channel. Our proposed codebook design for the \ac{MISO} beamforming case separately quantizes the channel direction and the residual phase.
The performance loss incurred by the finite rate feedback compared to perfect \ac{CSIT} is analyzed. Bounds of the power and capacity loss 
are derived.
Our work provides a path to making the assumption of CSIT in SISO and MISO systems with one-bit ADCs more realistic.

\section{System Model}


\begin{figure}[t]
\begin{centering}
\includegraphics[width = 1 \columnwidth]{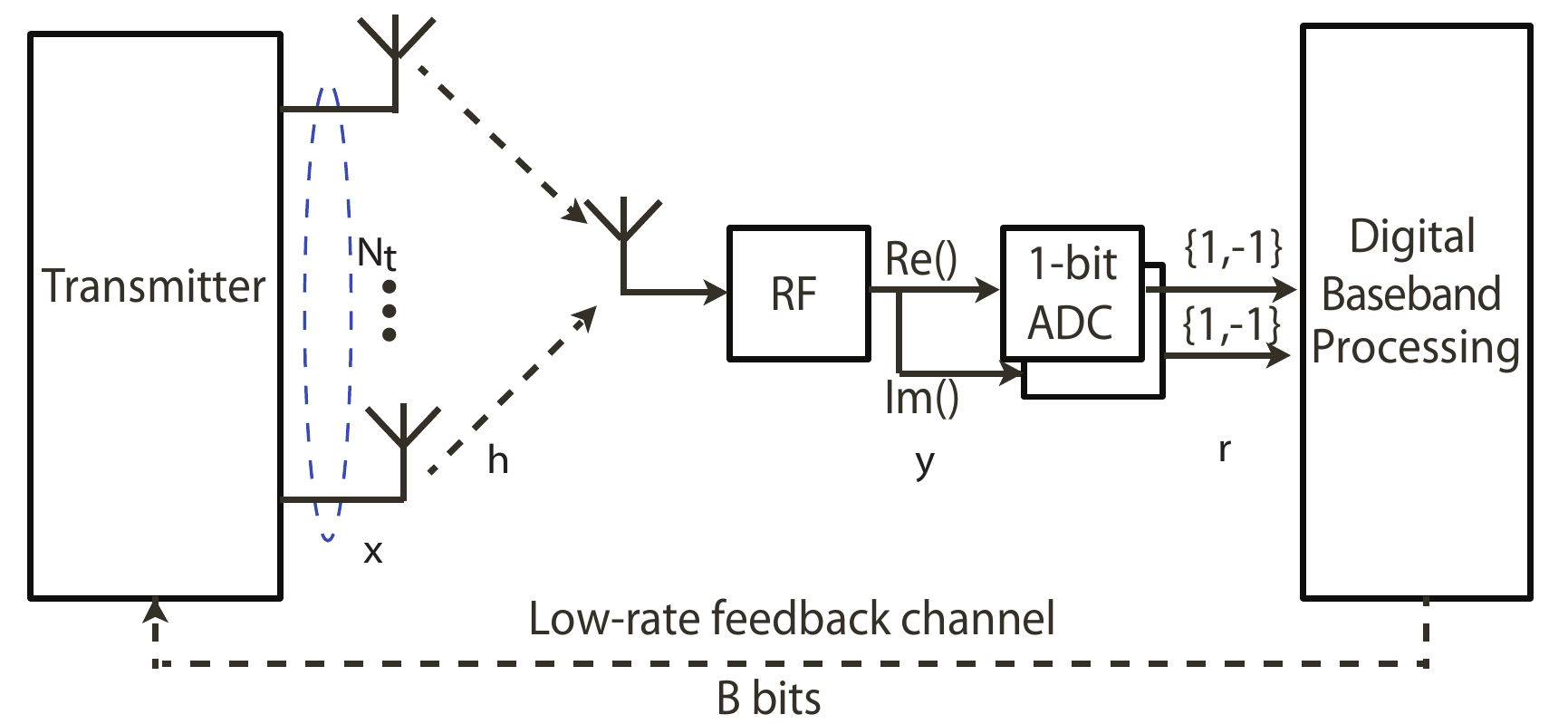}
\vspace{-0.1cm}
\centering
 \caption{A MISO system with one-bit quantization and limited feedback.  At the receiver side, there are two one-bit ADCs. There is also a low-rate feedback path from the receiver to the transmitter. Note that there is no limitation on the structure of the transmitter.}\label{fig:System_model}
\end{centering}
\vspace{-0.3cm}
\end{figure}

Consider a MISO system with one-bit quantization, as shown in Fig. \ref{fig:System_model}.
There are $\Nt$ antennas at the transmitter and single antenna at the receiver.
Assuming perfect synchronization and a narrowband channel, the baseband received signal in this MISO system is
\begin{equation}
  y = \mb{h}^*\mb{x}+ n
\end{equation}
where $\mb{h} \in \mathbb{C}^{\Nt\times 1}$ is the channel vector, $\mb{x}\in \mathbb{C}^{\Nt \times 1}$ is the signal sent by the transmitter, $y\in \mathbb{C}$ is the received signal before quantization, and $n \sim \mathcal{CN}(0, 1)$ is the circularly symmetric complex Gaussian noise. The average transmit power is $\Pt$, i.e., $\mathbb{E}[\bx^* \bx] = \Pt$.
In our system, there are two one-bit resolution quantizers that separately quantize the real and imaginary part of the received signal.
The output after the one-bit quantization is
\begin{equation}
  r = \mr{sgn}\left(y\right) = \mr{sgn}\left(\mb{h}^*\mb{x}+n\right),
\end{equation}
where $\mr{sgn}()$ is the signum function applied separately to the real and imaginary parts. Therefore, the quantization output $r \in \{1+\j, 1-\j, -1+\j, -1-\j\}$.

In this paper, we assume the receiver has perfect channel state information. This is justified by the prior work on channel estimation with one-bit quantization, for example \cite{Ivrlac_WSA07, Dabeer_ICC10, Mezghani_WSA10, Mezghani_WSA12, Mo_Jianhua_Asilomar14, Wen_Chao-Kai_arxiv15b, Jacobsson_arxiv15}.  Furthermore, the feedback is assumed to delay and error free, as is typical in limited feedback problems. Adding realism to the feedback channel is an interesting topic for future work.

\section{Quantized SISO Channel with Limited Feedback}

We first consider the SISO channel as a special case. Since  $\Nt=1$, the received signal is
\begin{eqnarray}
  r &=& \mr{sgn}\left( y \right) = \mr{sgn}\left( h x + n \right).
\end{eqnarray}
Denote the phase of $h$ as $\angle h$. As shown in our previous work \cite[Lemma 1]{Mo_Jianhua_TSP15}, the capacity-achieving scheme is to transmit rotated QPSK symbols, i.e.,
\begin{equation}
  \mr{Pr}\left[x = \sqrt{\Pt} e^{\j\left( \frac{k \pi}{2} + \frac{\pi}{4} - \angle h \right)}\right] = \frac{1}{4}, \; \mr{for} \; k=0, 1, 2 \;\mr{and}\; 3.
\end{equation}
The term $-\angle h$ in the transmitted signals is introduced to pre-cancel the phase rotation of the channel such that the receiver will observe a regular QPSK signal.

If $h$ is unknown at the transmitter, then only the phase needs to be quantized and fed back to the transmitter. Since the QPSK constellation is unchanged for a 90-degree rotation, only $\mr{mod} \left(\angle h, \frac{\pi}{2}\right)$ instead of $\angle h$ needs to be fed back. Now assume $B$ bits are used to uniformly quantize the region $[0, \frac{\pi}{2}]$. Uniform quantization is reasonable since for most statistical channel models the phase of the SISO channel is uniformly distributed. The codebook is then $\Phi=\{\phi_i = \frac{i\pi}{2^{B+1}} + \frac{\pi}{2^{B+2}}, 0 \leq i \leq 2^B-1 \}$. For instance, $\Phi=\{\frac{\pi}{8}, \frac{3\pi}{8}\}$ if $B=1$.
The receiver sends the index $i$ of $\hat{\phi}$ to the transmitter such that

\begin{eqnarray}
  \hat{\phi} = \underset{\phi_m \in \Phi}{\arg \min} \left|\mathrm{mod} \left(\angle h, \frac{\pi}{2} \right) - \phi_m \right|.
\end{eqnarray}
Based on the feedback index $i$, the transmitter sends rotated QPSK signals with uniform probabilities, i.e.,
\begin{equation}
  \mr{Pr}\left[x = \sqrt{\Pt} e^{\j\left( \frac{k \pi}{2} + \frac{\pi}{4} - \hat{\phi} \right)}\right] = \frac{1}{4}, \; \mr{for} \; k=0, 1, 2 \;\mr{and}\; 3.
\end{equation}
The received signal after quantization is
\begin{eqnarray}
  r = \mr{sgn} \left( |h| \sqrt{\Pt} e^{\j\left( \frac{k \pi}{2} + \frac{\pi}{4} - \hat{\phi} + \angle h \right)} + n \right).
\end{eqnarray}

The channel has four possible inputs and four possible outputs. This is a discrete-input discrete-output channel. Therefore the channel capacity is
\begin{eqnarray}
  & &C_{\mr{SISO}}^{\mr{fb}} \nonumber \\
  &=& 2 - \Hb \left(  Q\left( \sqrt{2 \Pt \left|h\right|^2 \sin^2 \left( \frac{\pi}{4}- \theta \right)}\right) \right) \nonumber \\
  & &  \quad - \Hb \left(  Q\left(\sqrt{2 \Pt  \left|h\right|^2 \cos^2 \left(\frac{\pi}{4} - \theta \right)}\right) \right) \\
  &=& 2 - \Hb \left(  Q\left(\sqrt{\Pt  \left|h\right|^2 \left(1 - \sin 2 \theta \right)}\right) \right) \nonumber \\
  & &  \quad - \Hb \left(  Q\left( \sqrt{ \Pt \left|h\right|^2 \left(1 + \sin 2 \theta \right)}\right) \right) ,
\end{eqnarray}
where $\theta := \hat{\phi} - \mathrm{mod} \left(\angle h, \frac{\pi}{2} \right)$ is the quantization error, $\Hb(p) := -p \log_2 p -(1-p) \log_2 (1-p)$ is the binary entropy function, and $Q(\cdot)$ is the tail probability of the standard normal distribution.

\begin{figure}[t]
\begin{centering}
\includegraphics[width = 1\columnwidth]{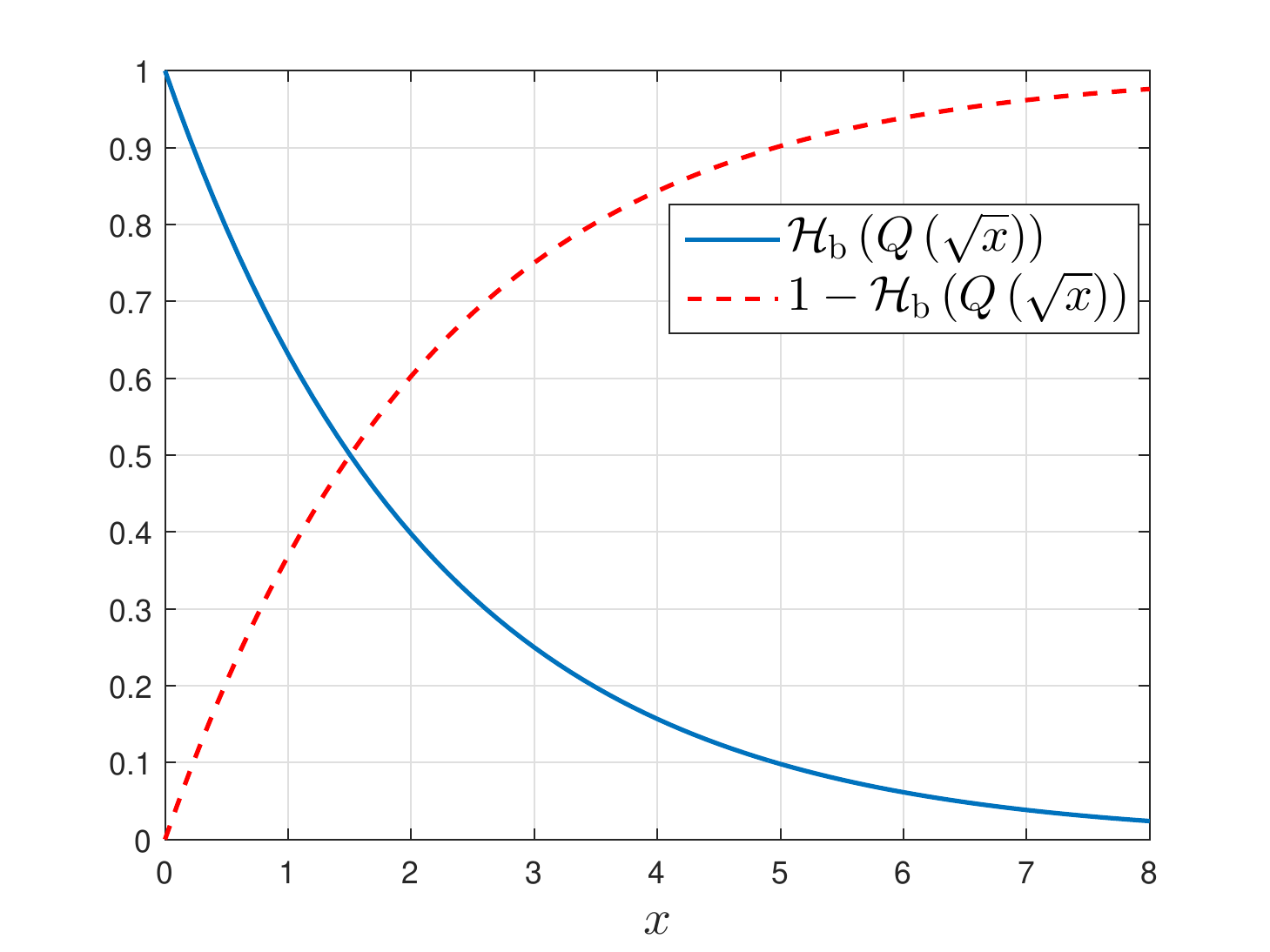}
\vspace{-0.1cm}
\centering
 \caption{The figure shows $\Hb \left(Q \left( \sqrt{x}\right) \right)$ and $1-\Hb \left(Q \left( \sqrt{x}\right) \right)$ versus $x$. It is seen that $\Hb \left(Q \left( \sqrt{x}\right) \right)$ is a decreasing and convex function of $x$. 
 	}\label{fig:H_b_Q_sqrt_x}
\end{centering}
\vspace{-0.3cm}
\end{figure}

In Fig. \ref{fig:H_b_Q_sqrt_x}, we plot the function $\Hb \left(Q \left( \sqrt{x}\right) \right)$.
Since $\Hb \left(Q \left( \sqrt{x}\right) \right)$ is decreasing with $x$, it follows that
\begin{eqnarray} \label{eq:Capacity_SISO_fb_lb}
  C_{\mr{SISO}}^{\mr{fb}} \geq 2 \left(1 - \Hb \left(  Q\left(\sqrt{\Pt  \left|h\right|^2 \left(1 - \sin 2 \left|\theta \right| \right)}\right) \right) \right).
\end{eqnarray}
The channel capacity with perfect \ac{CSIT} is \cite[Lemma 1]{Mo_Jianhua_TSP15}
\begin{equation} \label{eq:Capacity_SISO}
  C_{\mr{SISO}} = 2\left(1 - \Hb \left(  Q\left(\sqrt{\Pt \left|h\right|^2} \right) \right)\right).
\end{equation}

Comparing \eqref{eq:Capacity_SISO_fb_lb} and \eqref{eq:Capacity_SISO}, the power loss factor is $\frac{1}{1- \sin 2 \left| \theta \right|} $. We want to minimize the power loss, or equivalently maximize the term $1 - \sin 2 \left|\theta \right|$.
Since the quantization error $\theta \in \left[ -\frac{\pi}{2^{B+2}}, \frac{\pi}{2^{B+2}} \right]$ by the uniform quantization scheme, 
it follows that ${1- \sin 2 \left| \theta \right|} > {1- \sin \frac{\pi}{2^{B+1}}}$. When $B=1$, $1- \sin 2 |\theta| \geq 1 - \frac{1}{\sqrt{2}}$, which means that there is at most $5.33$ $\mr{dB}$ power loss. In addition, as $\theta$ is uniformly distributed, the average power loss is
\begin{eqnarray}
  & & \mathbb{E}_{\theta} \left[ 1 - \sin 2 |\theta| \right] \\
  &=& \frac{1}{\frac{2 \pi}{2^{B+2}}} \int_{-\frac{\pi}{2^{B+2}}}^{\frac{\pi}{2^{B+2}}} \left(1- \sin 2 |\theta| \right)  \mr{d} \theta \\
  &=& 1 - \frac{\sin^2 \left(\frac{\pi}{2^{B+2}}\right)}{\frac{\pi}{2^{B+2}}} \label{eq:phase_quan_bound_exact}\\
  &\stackrel{(a)}{>}& 1 - \frac{\pi}{2^{B+2}} \\
  &>& 1 - 2^{-B} \label{eq:phase_quan_bound_loose} ,
\end{eqnarray}
where $(a)$ follows from $\sin x < x$ for $0< x < \frac{\pi}{2}$.
Therefore, the average power loss is at most $3$ $\mr{dB}$ with only one bit feedback\footnote{A tighter bound is $2$ $\mr{dB}$ by evaluating \eqref{eq:phase_quan_bound_exact} with $B=1$.}.
In the simulation, we will show that with only one bit feedback, the performance is close to that with perfect \ac{CSIT}.

\section{Quantized MISO Channel with Limited Feedback}

Similar to the MISO system with infinite-resolution ADCs, random vector quantization (RVQ),
which performs close to the optimal quantization and is amenable to analysis \cite{Jindal_IT06, Au-Yeung_TWC07},
is adopted to quantize the direction of channel $\bh$. We assume that $B_1$ out of the total $B$ bits are used to convey the channel direction information. The codebook is $\mathcal{W} = \left\{\bv_0, \bv_1, \cdots, \bv_{2^{B_1}-1} \right\}$ where each of the quantization vectors is independently chosen from the isotropic distribution on the Grassmannian manifold $\mathcal{G}(\Nt, 1)$ \cite{Love_IT03}. The receiver sends back the index of $\bv$ maximizing $|\bh^* \bv|$.

Besides the channel direction information, the remaining $B_2 = B -B_1$ bits are used to quantized the phase of the equivalent channel, i.e., $\angle \left(\bh^* \bv \right)$ (denoted as \emph{residual phase} afterwards).
The second codebook quantizing the residual phase is $\Phi=\{\phi_i = \frac{i\pi}{2^{B+1}} + \frac{\pi}{2^{B+2}}, 0 \leq i \leq 2^{B_2}-1 \}$.
The receiver feeds back the index $i$ of $\hat{\phi}$ such that
\begin{eqnarray}
  \hat{\phi} = \underset{\phi_m}{\arg \min} \left| \mathrm{mod} \left(\angle \left( \bh^* \bv \right), \frac{\pi}{2} \right) - \phi_m \right|.
\end{eqnarray}

The transmitter adopts matched filter beamforming and QPSK signaling based on the feedback bits, i.e.,
\begin{equation}
  \mr{Pr}\left[\mb{x} =  \sqrt{\Pt} \bv e^{\j \left( \frac{k \pi}{2} + \frac{\pi}{4} - \hat{\phi} \right)}\right] = \frac{1}{4}, \; \mr{for} \; k=0, 1, 2 \; \mr{and} \; 3.
\end{equation}

The received signal after quantization is
\begin{eqnarray}
  r = \mr{sgn} \left( \sqrt{\Pt} \bh^*\bv e^{\j\left( \frac{k \pi}{2} + \frac{\pi}{4} - \hat{\phi} \right)} + n \right).
\end{eqnarray}

Similar to the SISO case, this channel is also a discrete-input discrete-output channel. The capacity is derived as
\begin{eqnarray} 
  & &C_{\mr{MISO}}^{\mr{fb}} \nonumber \\
  &=& 2 - \Hb \left(  Q\left(\sqrt{2 \Pt |\bh^* \bv |^2 \sin^2 \left( \frac{\pi}{4}- \theta \right)}\right) \right) \nonumber \\
  & &  \quad - \Hb \left(  Q\left( \sqrt{2 \Pt |\bh^* \bv |^2 \cos^2 \left( \frac{\pi}{4} - \theta \right)}\right) \right) \\
  &=& 2 - \Hb \left(  Q\left(\sqrt{\Pt \|\bh \|^2 \cos^2 \beta \left(1 - \sin 2 \theta \right)}\right) \right) \nonumber \\
  & &  \quad - \Hb \left(  Q\left(\sqrt{ \Pt \|\bh \|^2 \cos^2 \beta \left(1 + \sin 2 \theta \right)}\right) \right) ,
\end{eqnarray}
where $\theta := \hat{\phi} - \mr{mod} \left(\angle \left( \bh^* \bv \right), \frac{\pi}{2}\right)$ and $\cos \beta := \frac{\left|\bh^* \bv \right|}{ \| \bh\|}$.
A lower bound of the capacity is
\begin{eqnarray} \label{eq:Capacity_MISO_fb_lb}
  & & C_{\mr{MISO}}^{\mr{fb}} \nonumber \\
  &\geq & 2 \left(1 - \Hb \left(  Q\left( \sqrt{\Pt \|\bh \|^2 \cos^2 \beta \left(1 - \sin 2 |\theta| \right)}\right) \right) \right). \nonumber \\
\end{eqnarray}

The channel capacity with perfect \ac{CSIT}, derived in \cite{Mo_Jianhua_TSP15}, is
\begin{equation} \label{eq:Capacity_MISO}
  C_{\mr{MISO}} = 2\left(1 - \Hb\left(  Q\left( \sqrt{\Pt \|\mb{h}\|^2}\right) \right)\right).
\end{equation}

Comparing \eqref{eq:Capacity_MISO_fb_lb} and \eqref{eq:Capacity_MISO}, we want to maximize the term ${\cos^2 \beta \left( 1 - \sin 2 |\theta| \right)}$ to minimize the power loss.
Averaging over the codebook $\mathcal{W}$ and the residual phase $\theta$, we have
\begin{eqnarray}
  & &\mathbb{E}_{\mathcal{W}, \theta} \left[ \cos^2 \beta \left( 1 - \sin 2|\theta|\right) \right] \\
  &\stackrel{(a)}{=}&\mathbb{E}_{\mathcal{W}} \left[ \cos^2 \beta \right] \mathbb{E}_{\theta} \left[ 1 - \sin 2|\theta| \right] \\
  &\stackrel{(b)}{\geq}& \left( 1- 2^{-\frac{B_1}{\Nt-1}}\right)  \left( 1- {2^{-B_2}} \right) \label{eq:MISO_power_loss}
\end{eqnarray}
where $(a)$ is by noting that $|\bh^*\bv|$ and $\angle \left( \bh^* \bv\right)$ are independent for RVQ,
$(b)$ follows from the facts $\mathbb{E}_{\mathcal{W}} \left[\cos^2 \beta \right] > 1 - 2^{ -\frac{B_1}{\Nt-1}}$ \cite[Lemma 1]{Jindal_IT06} and $\mathbb{E}_{\theta} \left[ 1 - \sin 2 |\theta| \right] > 1 - 2^{-B_2}$ proved in \eqref{eq:phase_quan_bound_loose}.
From \eqref{eq:MISO_power_loss}, it is seen that when $B_1=\Nt-1$ and $B_2=1$, the average power loss is at most $6$ $\mr{dB}$.

Another performance metric is capacity loss, which may be more important than power loss. We now analyze the capacity loss caused by limited feedback. Note that the capacity of the quantized system saturates to $2$ $\mr{bps/Hz}$ at high SNR, which is a difference from the unquantized systems. The capacity loss incurred by finite-rate feedback is
\begin{eqnarray}
  C_{\mr{loss}} &=& C_{\mr{MISO}} - C_{\mr{MISO}}^{\mr{fb}}  \\
  &\leq& 2 - C_{\mr{MISO}}^{\mr{fb}} \\
  &\leq& 2 \Hb \left(  Q\left( \sqrt{\Pt \|\bh \|^2 \cos^2 \beta \left(1 - \sin 2 |\theta| \right)}\right) \right).
\end{eqnarray}

To ensure $C_{\mr{loss}} \leq 2\epsilon$, we required
\begin{eqnarray}
  \Pt \|\bh \|^2 \cos^2 \beta \left(1 - \sin 2 |\theta| \right) \geq \delta,
\end{eqnarray}
where $\Hb \left(  Q\left( \sqrt{\delta}\right) \right) = \epsilon$. Plugging in \eqref{eq:MISO_power_loss} and assuming $\mathbb{E}\left[\| \bh \|^2\right] = \Nt$, we obtain
\begin{eqnarray} \label{eq:MISO_high_SNR_fb}
  \left( 1- 2^{-\frac{B_1}{\Nt-1}}\right)  \left( 1- {2^{-B_2}} \right) \geq \frac{\delta}{\Pt \Nt}.
\end{eqnarray}

In \eqref{eq:MISO_high_SNR_fb}, we see that given fixed capacity loss, the required number of feedback bits actually {\it decreases} with the transmit signal power. This is in striking contrast with the unquantized MISO systems.

In Fig. \ref{fig:H_b_Q_sqrt_x}, it is shown that $\Hb \left( Q \left( \sqrt{5} \right) \right) \approx 0.1$. Therefore, if the numbers of feedback bits satisfy
\begin{eqnarray} \label{eq:MISO_high_SNR_fb_example}
  \left( 1- 2^{-\frac{B_1}{\Nt-1}}\right)  \left( 1- {2^{-B_2}} \right) \geq \frac{5}{\Pt \Nt},
\end{eqnarray}
then the capacity loss is less than $0.2$ $\mr{bps/Hz}$, or equivalently $90\%$ of the upper bound (which is $2$ $\mr{bps/Hz}$) is achieved.

\section{Simulation Results}

In this section, we evaluate the performance of the proposed limited feedback scheme. We compute the capacity for each channel realization then averaged over 1000 channel realizations with Rayleigh fading, i.e., $\bh \sim \mathcal{CN}(\mathbf{0}, \mathbf{I}_{\Nt})$. In the figures, $\mathrm{SNR (dB)} \triangleq 10 \log_{10} \Pt$ since the noise is assumed to have unit variance.

\begin{figure}[t]
\begin{centering}
\includegraphics[width=0.95\columnwidth]{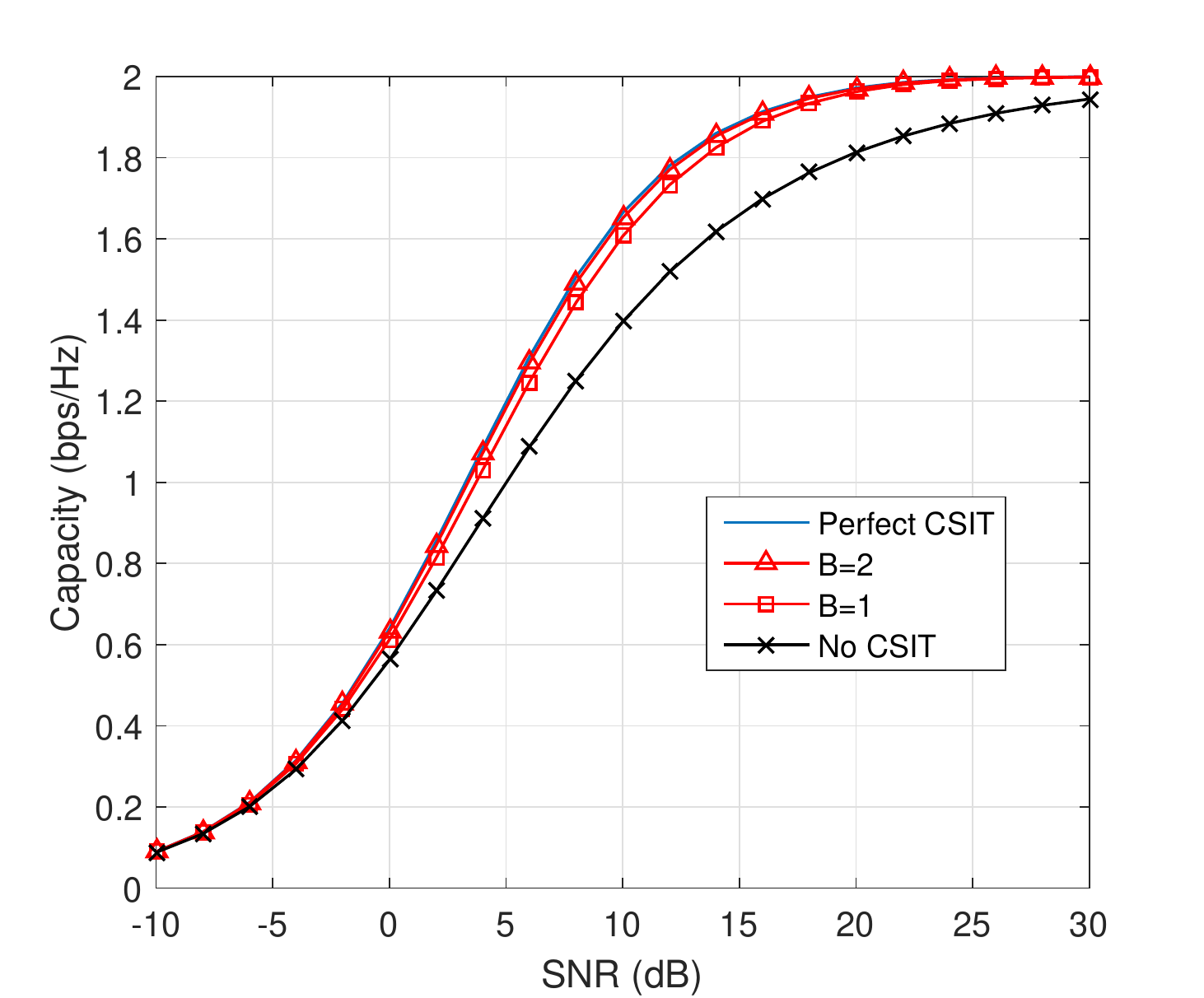}
\vspace{-0.1cm}
\centering
 \caption{The capacity of a SISO system with CSIT, no CSIT and feedback.
 }\label{fig:SISO_Capacity}
\end{centering}
\vspace{-0.3cm}
\end{figure}

In Fig. \ref{fig:SISO_Capacity}, we  compare the capacities of SISO channel with perfect \ac{CSIT}, limited feedback and no CSIT.  As shown in the figure, the capacity with two bits feedback is almost same to that with perfect CSIT. In addition, even with single bit feedback, the power loss is very small, i.e., less than 1 $\mr{dB}$. Without CSIT, the capacity loss is much larger, especially at high SNR. Taking into account both the feedback overhead and the capacity loss, it is reasonable to set $B=1$ in practice.

\begin{figure}[t]
\begin{centering}
\includegraphics[width=0.95\columnwidth]{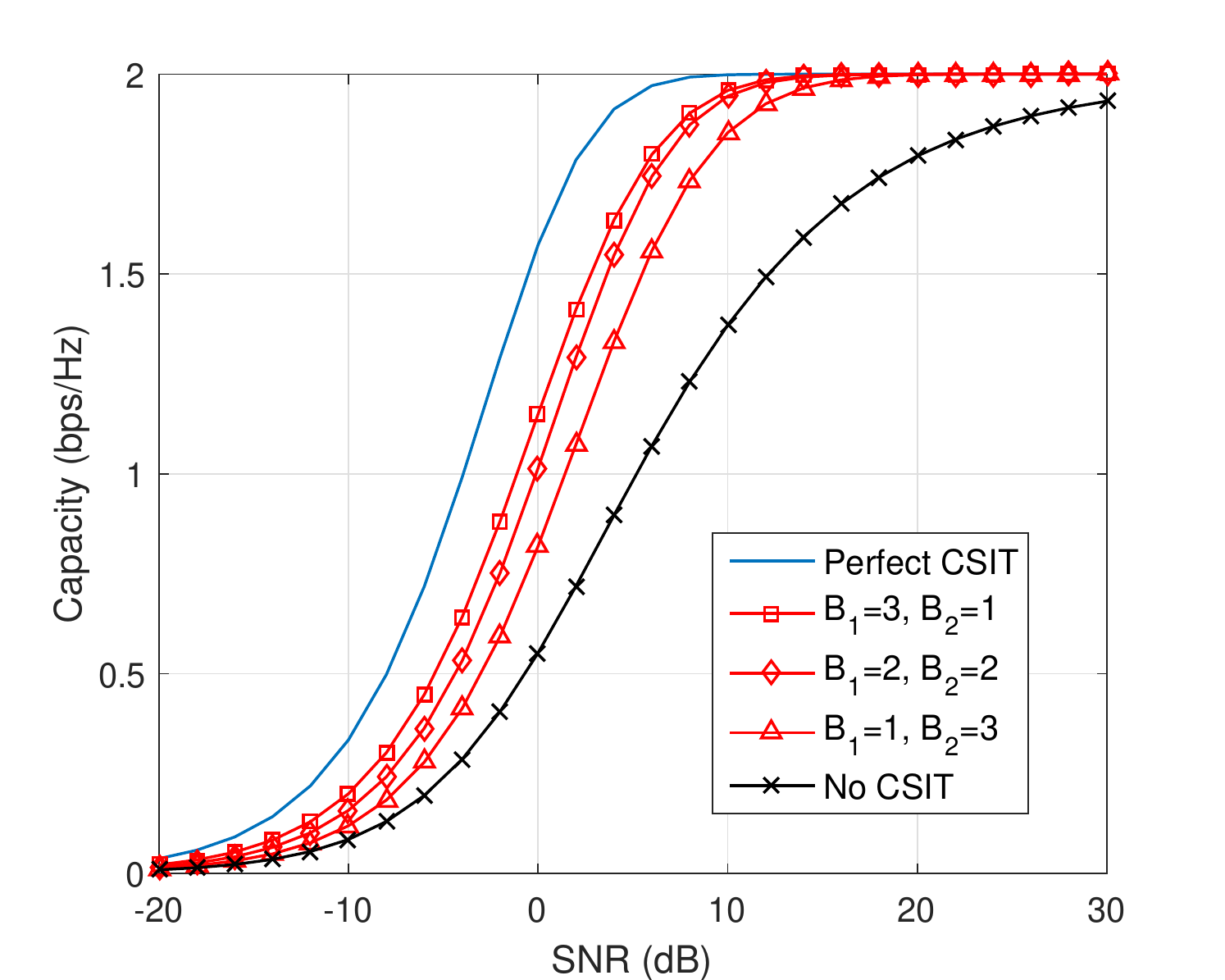}
\vspace{-0.1cm}
\centering
 \caption{The capacity of a MISO system with CSIT, no CSIT and limited feedback when $\Nt=4$.
 }\label{fig:MISO_Capacity_Nt_4}
\end{centering}
\vspace{-0.3cm}
\end{figure}

\begin{figure}[t]
\begin{centering}
\includegraphics[width=0.95\columnwidth]{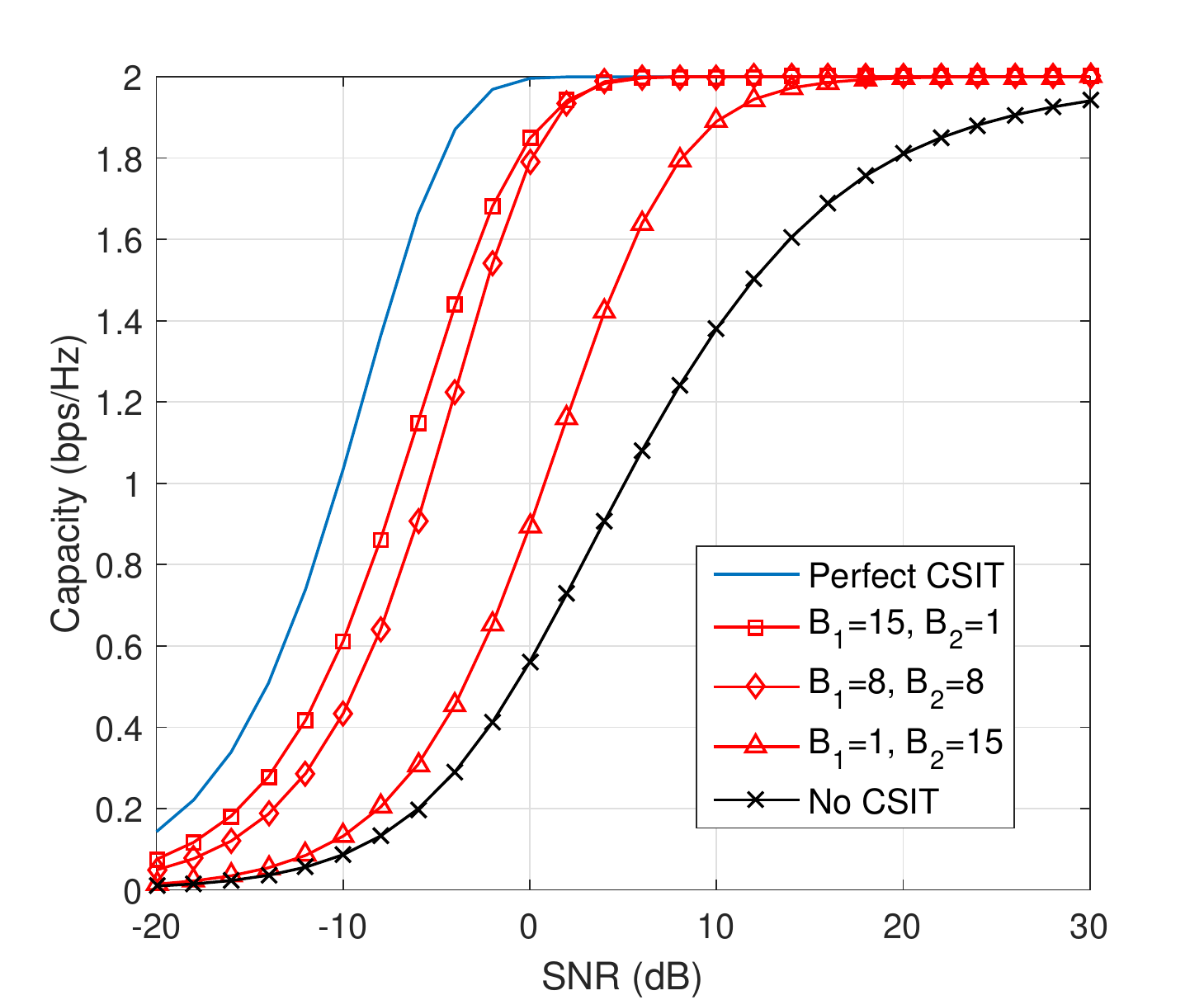}
\vspace{-0.1cm}
\centering
 \caption{The capacity of a MISO system with CSIT, no CSIT and limited feedback when $\Nt=16$.
 }\label{fig:MISO_Capacity_Nt_16}
\end{centering}
\vspace{-0.3cm}
\end{figure}

\begin{figure}[t]
\begin{centering}
\includegraphics[width=0.95\columnwidth]{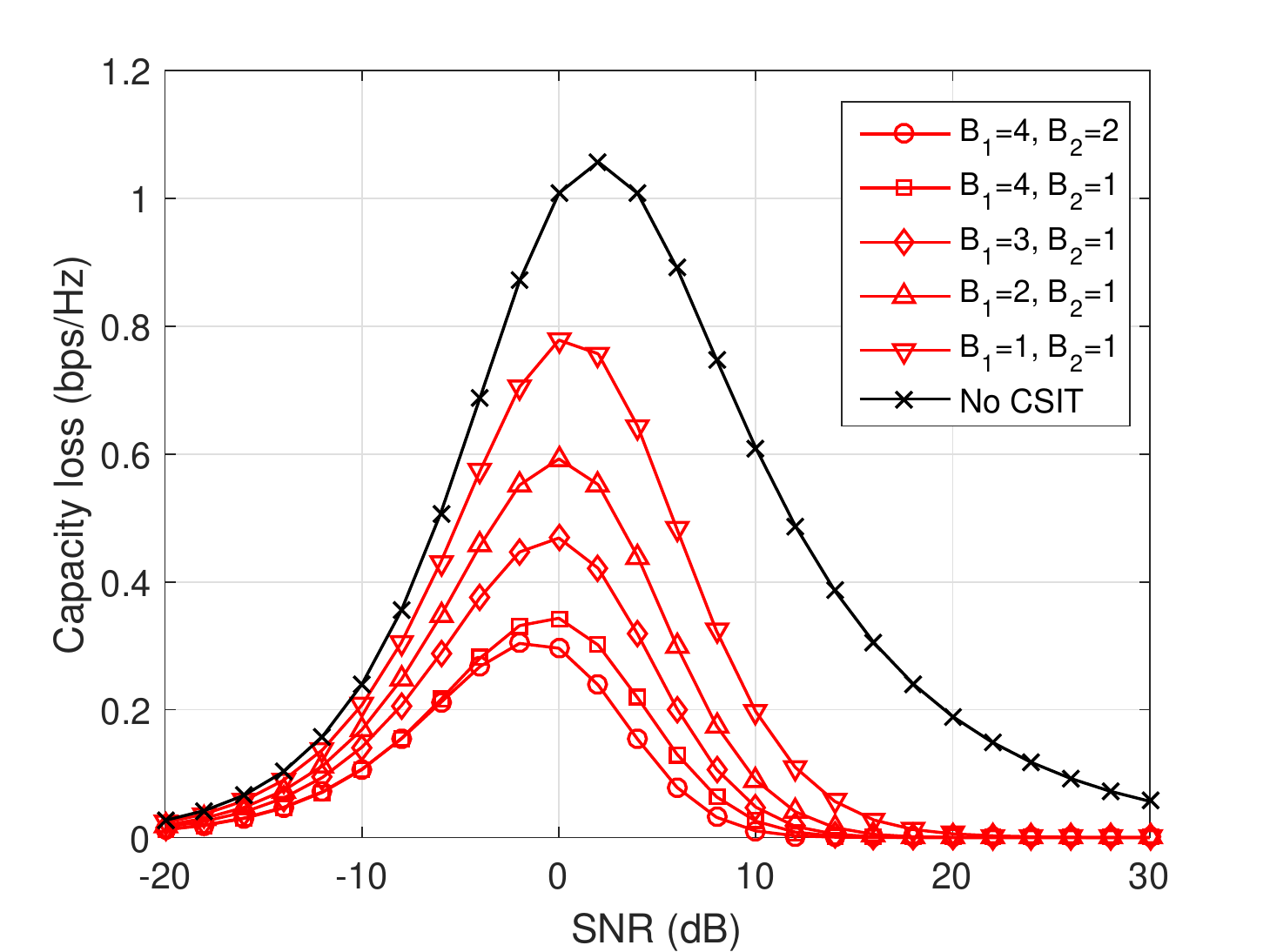}
\vspace{-0.1cm}
\centering
 \caption{The capacity loss $C_{\mr{loss}} = C_{\mr{MISO}} - C_{\mr{MISO}}^{\mr{fb}}$ incurred by finite-rate feedback in a MISO system with $\Nt=4$.
 }\label{fig:Capacity_loss_Nt_4}
\end{centering}
\vspace{-0.3cm}
\end{figure}

Figs. \ref{fig:MISO_Capacity_Nt_4} - \ref{fig:Capacity_loss_Nt_4} show the performance of the proposed limited feedback scheme in MISO channel. The entries of channel $\bh$ follow the IID circularly symmetric complex Gaussian distribution with unit variance.

In \figref{fig:MISO_Capacity_Nt_4}, we plot the capacities of MISO with four antennas and four bits feedback. Three different allocations of the feedback bits are compared. It is found that the case `$B_1=3, B_2=1$' has the best performance with power loss around $3$ $\mr{dB}$, which is consistent with our analysis in \eqref{eq:MISO_power_loss} which states the power loss is upper bounded by $6$ $\mr{dB}$. In Fig. \ref{fig:MISO_Capacity_Nt_16}, we show another example with 16 antennas and 16 bits feedback. It is shown that the case `$B_1=15, B_2=1$' is the best one. We therefore conclude that more bits should be assigned to feed back the channel direction information.

In \figref{fig:Capacity_loss_Nt_4}, we show the capacity loss for different values of $B_1$ and $B_2$. As expected, the capacity loss decreases as $B_1$ and $B_2$ increase. We also find that at high SNR, the capacity losses incurred by limited feedback converge to zero. For instance, when the transmitter power is larger than $11$ $\mr{dB}$, even with $B_1=B_2=1$, the capacity loss is less than $0.2$ $\mr{bps/Hz}$, which implies that the capacity with only two bit feedback achieves $90\%$ of the capacity with perfect CSIT. The result verifies our analysis in \eqref{eq:MISO_high_SNR_fb_example}.
Note that the capacity loss at low SNR is also small since $C_{\mr{MISO}}$ is small.

%
%

\section{Conclusions}

In this paper, we developed an approach for limited feedback in SISO and MISO channels with one-bit ADCs. For the SISO channel, only the phase of the channel is quantized while in the MISO channel, the channel direction and residual phase are both quantized and fed back to the transmitter. We evaluated the power and capacity losses incurred by the use of limited feedback. Based on our analyses and simulation results, we made two important observations. First, feeding back only one bit for the phase (or residual phase in MISO channel) is enough to guarantee good performance in the examples considered. Second, when the capacity of the quantized channel is saturated, the required number of feedback bits guaranteeing a small capacity loss decreases with SNR.

There are several potential directions for future work. Our numerical results were based on the IID Gaussian channel with small numbers of antennas. In mmWave systems - a promising application of one-bit ADCs - the channels will likely be correlated depending on the number of scattering clusters and the angle spread. It would be interesting to develop techniques that also work for large correlated channels. A natural extension of our work would be to MIMO communication channels. This is complicated by the complicated structure of the capacity-optimum signaling distribution and the potential for different choices of precoders (see \cite[Section IV]{Mo_Jianhua_TSP15}). 
Another possible direction is to combine the two separate stages, channel estimation and limited feedback, together. In this case, the feedback bits are decided directly by the ADC outputs instead of the estimated CSI at the receiver.

\bibliographystyle{IEEEtran}
\bibliography{IEEEabrv,One_bit_quantization}
\end{document}